\documentclass[iop]{emulateapj}
\usepackage{color}

\makeatletter
\def\blfootnote{\xdef\@thefnmark{}\@footnotetext}
\makeatother

\begin{document}
\shorttitle{Accretion disk radiation} \shortauthors{Koutsantoniou
\& Contopoulos}

\def\gsim{\mathrel{\raise.5ex\hbox{$>$}\mkern-14mu
             \lower_{\rm 0}.6ex\hbox{$\sim$}}}
\def\lsim{\mathrel{\raise.3ex\hbox{$<$}\mkern-14mu
             \lower_{\rm 0}.6ex\hbox{$\sim$}}}

\author{
Leela E.
Koutsantoniou\altaffilmark{1,}\altaffilmark{2,}\altaffilmark{*}
and Ioannis Contopoulos\altaffilmark{1,}\altaffilmark{$\dagger$}} \affil{\altaffilmark{1}
Research Center for Astronomy and Applied Mathematics, Academy of
Athens, Athens
11527, Greece \\  \\
\altaffilmark{2}Department of Astrophysics, Astronomy and
Mechanics,
Faculty of Physics, University of Athens, \\
Panepistimiopolis Zografos, Athens 15784, Greece}
\blfootnote{* leelamk@phys.uoa.gr}
\blfootnote{$\dagger$ icontop@academyofathens.gr}

\title{Accretion disk radiation dynamics and the cosmic battery}



\begin{abstract}We investigate the dynamics of radiation in
the surface layers of an optically thick astrophysical accretion
disk around a Kerr black hole. The source of the radiation is the
surface of the accretion disk itself, and not a central object as
in previous studies of the Poynting-Robertson effect. We generate
numerical sky maps from photon trajectories that originate on the
surface of the disk as seen from the inner edge of the disk at the
position of the innermost stable circular orbit (ISCO). We
investigate several accretion disk morphologies with a
Shakura-Sunyaev surface temperature distribution. Finally, we
calculate the electromotive source of the Cosmic Battery mechanism
around the inner edge of the accretion disk and obtain
characteristic timescales for the generation of astrophysical
magnetic fields.
\end{abstract}



\section{Introduction}

From a distance, astrophysical sources of luminosity $L$ look like
point sources of isotropic radiation. In that limit, radiation
introduces a radial force component per proton
\begin{equation}
f_{\rm rad}^r=\frac{L \sigma_T}{4\pi r^2 c}\ , \label{fradNR}
\end{equation}
which obviously modifies the dynamics of the surrounding plasma.
Here, $\sigma_T$ is the Thomson cross-section
for photon scattering by the plasma electrons, and $r$ is the
distance of the orbiting proton. It is important to realize that
radiation pressure is felt only by the plasma electrons. As we
will show in detail below, the protons feel the radiation force
through a radial electric field that develops between them and the
electrons.

Eq.~(\ref{fradNR}) is not the only force component in an isotropic
radiation field. Matter is in orbit around the center, and
therefore, one needs to also take into account the
Poynting-Robertson radiation drag. This effect (hereafter PR) was
first described in 1903 by J. H. Poynting, but was revisited and
fully explained much later in 1937 by H. P. Robertson, with a
significant contribution from J. Larmor. Its main application has
been in the study of orbits of particles (mainly dust grains)
around the Sun. Without radiation drag, the orbits are circular or
elliptical. In the presence of radiation, the grains scatter
photons emitted by the central star, lose angular momentum, and
inspiral toward the star. The question that arises, is how can a
force perpendicular to the motion slow down the grain. The answer
is obvious when we view the radiation field in the frame of the
moving particle. In that frame, the radiation field is aberrated,
namely slightly concentrated or beamed in the direction of motion.
This abberation results in an extra radiation force opposite to
the target's motion that slows it down. In flat spacetime, the PR
azimuthal drag force per proton
can be directly calculated as
\begin{equation}
f_{\rm PR}^\phi=-f_{\rm rad}^r \frac{{\rm v}^\phi}{c}=-\frac{L
\sigma_T{\rm v}^\phi}{4\pi r^2 c^2}\ . \label{PRSR}
\end{equation}
Here, ${\rm v}^\phi$ is the plasma azimuthal velocity. Another way
to view the same effect is through the equivalence of energy and
mass, and the conservation of momentum. If we could separate the
scattering process into absorption and re-emission of the incoming
photon by the target we would see that as the photon is absorbed,
the `mass' of the target increases, and therefore its speed
decreases. During the re-emission, however, the outgoing photon
carries away mass as well as momentum, leaving the target's speed
unchanged. In total, every time the target `absorbs' a photon, it
slows down and falls into a lower orbit. This is the reason why
the PR effect is called a drag or braking. As before, the protons
in a plasma feel the PR drag through an azimuthal electric field
that develops between them and the electrons.

The astrophysical setting of an accreting rotating black hole is
far more complex, and the dynamics of its radiation field have not
been thoroughly investigated before. In Kerr spacetime, photon
orbits are curved and there are several shifts that change the
photon frequency during its travel through the curved and
inhomogeneous spacetime. The fact that now the source of radiation
is also rotating, further complicates the problem by introducing
an extra Doppler shift. If the central object is not a black hole
but a slowly rotating compact spherical star, then simplifications
can be made to the problem, making the process of solution
somewhat easier. This case was studied by Abramowicz, Ellis \&
Lanza~(1990), and in a series of papers by Miller and Lamb (Miller
\& Lamb~1993; Lamb \& Miller~1995; Miller \& Lamb~1996). Bini {\em
et al.}~(2008) \& (2011) studied the PR effect for photons and
test particles in the equatorial plane. The background spacetime
was taken to be Schwarzschild and Kerr. Oh {\em et al.}~(2010)
further studied `suspension orbits' around a slowly rotating
compact star with an isotropic radiation field. In all previous
works, a simple radiation field was assumed (one emanating from a
central spherical object). On the other hand, astrophysical
sources associated with accretion disks are expected to generate a
complex radiation field with ensuing complex dynamics in the
vicinity of the central object.

We were, therefore, compelled to develop a fully relativistic ray
tracing code that calculates the radiation field and takes into
account the spatial extent and rotation of its source. Our primary
interest is to study the dynamical effect of radiation around
rapidly rotating black holes, where extremely energetic photons
and strong magnetic fields are expected to be present. As material
is pulled from its surroundings (companion star, stellar wind,
interstellar medium, etc.) it forms a rotating disk around the
center. This material slowly loses angular momentum through a
number of different mechanisms of varying efficiency
(magnetorotational instability-MRI, magnetic braking by disk
winds/jets, PR drag, viscosity, turbulence, tidal forces,
gravitational waves, etc.) and inspirals towards the center. As it
accretes, the temperature rises to values of the order of $10^7$~K
or even higher. At these temperatures, matter radiates in the
X-ray part of the spectrum, giving us what we observe and know as
an X-ray binary (in the case of a stellar mass black hole), or an
AGN (in the case of a supermassive black hole). Since at this
point the disk speed is a considerable fraction of the speed of
light and the radiation luminosity a considerable fraction of its
Eddington value, we expect that the dynamical effects of radiation
will not only be noticeable but in some cases they may even be
dominant.

In the present paper, we calculate the radiation pressure at the
position of the Innermost Stable Circular Orbit (hereafter ISCO)
around a rotating Kerr black hole. Our work is original in that
\begin{enumerate}
\item[(a)] the source of radiation is the surface of the accretion
disk around the black hole which radiates as a black body of
various temperatures, and \item[(b)] we consider the effect on the
whole plasma and not on individual particles like dust grains (as
in the original discussion of the PR drag effect).
\end{enumerate}
In \S~2 we formulate the problem in general relativity and
describe our course of action toward its solution. In \S~3 we
present our problem setup and numerical results. In \S~4 we
discuss the effect of radiation pressure in generating
astrophysical magnetic fields through the mechanism of the Cosmic
Battery. Finally, in \S~5 we present our conclusions and discuss
prospects for future work.

\section{Mathematical formulation}

We will assume that the immediate environment of an accreting
rotating black hole can be described by the Kerr spacetime metric. We will follow closely the train of thought and logic of \citet{ML96} correcting minor mistakes along the way.
We will hereafter use geometrical units in
which ${c=G=1}$. Latin/Greek indices will denote space/spacetime
components respectively. We will also assume the Einstein notation
for summation over double indices.

\subsection{The Kerr metric}

A Kerr black hole is characterized by its mass $M$ and its angular
momentum $J$, or equivalently its spin parameter $a\equiv J/M$.
$a$ takes values between zero (for a non-rotating Schwarzschild
black hole) and $M$ (for a maximally rotating black hole).
Physical quantities are measured by Zero Angular Momentum
Observers (ZAMOs; also known as local Fiducial Observers or Fidos)
in their Locally Non-Rotating Frame (LNRF). In Boyer-Lindquist
$(t,r,\theta,\phi)$ coordinates, the Kerr metric reads
\begin{equation}
{\rm d}s^2 = - \alpha^2 {\rm d}t^2 + \varpi^2({\rm d}\phi -\omega
{\rm d}t)^2+\frac{\Sigma}{\Delta}{\rm d}r^2 +\Sigma {\rm
d}\theta^2\ ,
\end{equation}
where,
\[
\alpha = (\Delta \Sigma / A)^{1/2},
\]
\[
\omega = 2 a   M r /  A\ ,
\]
\[
\varpi =\left( A / \Sigma \right)^{1/2} \sin\theta\ .
\]
\[
\Sigma = r^2 + a^2 \cos^2\theta\ ,
\]
\[
\Delta = r^2 - 2 M r + a^2\ ,
\]
\[
A = (r^2 + a^2)^2 - a^2 \Delta \sin^2\theta
\]
Here,  $\alpha$ is the lapse function, $\omega$ is the angular
velocity of ZAMOs, and $\varpi$ is the cylindrical radius. It is
possible to transform four-vectors $p^\mu$ from the
Boyer-Lindquist coordinates to $p^{\hat{\mu}}$ in the LNRF and
back via the relations
\begin{equation}
p^\mu=e^\mu_{\hat{\nu}}p^{\hat{\nu}}\ \mbox{and}\
p^{\hat{\mu}}=e^{\hat{\mu}}_\nu p^\nu\ ,
\end{equation}
where the transformation tensor components are given explicitly as
\[
e^{t}_{\hat{t}}=\alpha^{-1}\ ,\
e^{\phi}_{\hat{t}}=\omega\alpha^{-1}\ ,
\]
\begin{equation}
e^{r}_{\hat{r}}=\left(\frac{\Sigma}{\Delta}\right)^{-1/2}\ ,\
e^{\theta}_{\hat{\theta}}=\Sigma^{-1/2}\ ,\
e^{\phi}_{\hat{\phi}}=\varpi^{-1}\ ,
\end{equation}
\[
e^{\hat{t}}_t=\alpha\ ,\ e^{\hat{\phi}}_t=-\omega\varpi\ ,
\]
\begin{equation}
e^{\hat{r}}_r=\left(\frac{\Sigma}{\Delta}\right)^{1/2}\ ,\
e^{\hat{\theta}}_\theta=\Sigma^{1/2}\ ,\
e^{\hat{\phi}}_\phi=\varpi\ .
\end{equation}
We have introduced here the notation that unhatted indices refer
to the Boyer-Lindquist frame whereas hatted ones to the LNRF.

We remind the reader that there are two characteristic surfaces
around a Kerr black hole. The first surface, the black hole event
horizon, corresponds to the outer root of the equation $\Delta=0$,
at $r_{\rm BH}=M+\sqrt{M^2-a^2}$. This is a sphere with radius
$r_{\rm BH}=2M$ for a non-rotating (Schwarzschild) black hole and
radius $M$ for a maximally rotating one. The event horizon is a
one-way membrane in the sense that particles, massive or massless,
that follow timelike or null geodesics respectively, can only
cross this surface in one direction: inwards. The second surface,
the static limit, corresponds to the surface inside which the
metric time component $g_{tt}\equiv \omega^2\varpi^2-\alpha^2$
becomes positive. The static limit and the event horizon are
always in contact at the rotation axis and differ the most on the
equatorial plane, where the static limit extends out to a radial
distance of $2M$ for all spin parameters. The region between the
horizon and the static limit is called the ergosphere. Another
characteristic radius is the position of the ISCO, $r_{\rm ISCO}$,
on the equator. It is generally assumed that the inner edge of the
accretion disk around the black hole coincides with the ISCO,
although this may vary considerably around exceedingly bright or
strongly magnetized sources (Balbus~2012, Contopoulos \&
Papadopoulos~2013). For prograde rotation, $r_{\rm ISCO}$ ranges
from $6M$ for a Schwarzschild black hole, to $M$ for a maximally
rotating one\footnote{We note that even though it seems like the
horizon, the ISCO and other surfaces coincide at $r=M$ for a
maximally rotating BH ($a=M$), they differ in proper radial
distances. This illusion is the result of a coordinate singularity
in the Boyer-Lindquist frame.}.

\subsection{The radiation force}

The radiation field in the deep interior of an optically thick
astrophysical disk is thermal. In the present paper, though, we
are mainly interested in the dynamics of radiation in the surface
layers which `see' photons coming from the whole surface of the
disk. The radiation force per proton, $f^i_{\rm rad}$, is a
non-gravitational term that enters the relativistic equation of
motion for the spatial velocity components
as
\begin{equation}
\frac{{\rm d}^2u^i}{{\rm d}\tau^2} + \Gamma^i_{\nu\kappa}u^\nu
u^\kappa = \frac{f^i_{\rm rad}}{m_{\rm p}}\ . \label{eqmotion}
\end{equation}
Here, $u^\mu$ is the plasma four-velocity, $\tau$ is the plasma
proper time, $\Gamma^\mu_{\nu\kappa}$ are the metric Christoffel
symbols, and $m_{\rm p}$ is the proton mass. We have assumed for
simplicity that the innermost accretion disk plasma consists of
protons and electrons. We have also assumed that electromagnetic
and gas pressure forces are negligible. The $\phi$-component of
the radiation force $f^\phi_{\rm rad}$ is the generalization of
the PR drag per proton.

A short comment on eq.~(\ref{eqmotion}) is in order here. The
radiation force is felt only by the plasma electrons (the Thomson
cross-section for the protons is roughly four million times
smaller than that for the electrons), yet the full plasma feels
the radiation force. This becomes clear when we consider the
equations of motion for the plasma electrons and protons
independently, namely
\begin{equation}
m_{\rm e}\frac{{\rm d}^2 u_{\rm e}^i}{{\rm d}\tau^2} + m_{\rm
e}\Gamma^i_{\nu\kappa}u_{\rm e}^\nu u_{\rm e}^\kappa = f^i_{\rm
rad}-{\rm e}E^i\ \mbox{and}\label{eqmotione}
\end{equation}
\begin{equation}
m_{\rm p}\frac{{\rm d}^2 u_{\rm p}^i}{{\rm d}\tau^2} + m_{\rm
p}\Gamma^i_{\nu\kappa}u_{\rm p}^\nu u_{\rm p}^\kappa = {\rm e}E^i\
,\label{eqmotioni}
\end{equation}
where the electron charge is equal to $-{\rm e}$, and $E^i$ is
called the `impressed' electromotive field that summarizes the
effect of the various non-electrical non-gravitational forces in
the induction equation (see below; Biermann \& Schluter~1951).
Since, $m_{\rm e}\ll m_{\rm p}$, one can ignore the $m_{\rm
e}$-terms in eq.~(\ref{eqmotione}) and obtain
\begin{equation}
E^i \approx \frac{f^i_{\rm rad}}{\rm e}\ .\label{E}
\end{equation}
Putting this back in eq.~(\ref{eqmotioni}) we obtain
\begin{equation}
\frac{{\rm d}^2 u_{\rm p}^i}{{\rm d}\tau^2} +
\Gamma^i_{\nu\kappa}u_{\rm p}^\nu u_{\rm p}^\kappa\approx
\frac{f^i_{\rm rad}}{m_{\rm p}}\ ,\label{eqmotioni2}
\end{equation}
which is just eq.~(\ref{eqmotion}) under the approximation that
$u_{\rm p}^\mu \approx u^\mu$. In other words, {\em the protons
feel the radiation force through the electric field that develops
between them and the electrons}. In \S~4 we will come back to this
important effect which has been discussed since the late 1940s,
but is often ignored by the younger generation of researchers.

$f^i_{\rm rad}$ is connected to the radiation flux $F^i$ through
the formula
\begin{equation}
f^i_{\rm rad}=\sigma_T F^i\ . \label{force}
\end{equation}
We have assumed for simplicity that the Thomson cross-section
$\sigma_T$ is independent of the radiation frequency and
scattering angle\footnote{For an incoming photon of energy 1 keV,
the cross section deviation from the uniform case of $\sigma_T$ is
in total very small and at most equal to $\sigma_T /2$. Therefore
we choose to ignore it in the present work, in order to avoid the
notably more complex calculations required in this case.}. The
radiation flux components $F^i$ are given by
\begin{equation}
F^i=h^i_\nu T^{\kappa \nu} u_\kappa\ ,
\end{equation}
where $T^{\kappa \nu}$ is the radiation stress-energy tensor, and
$h^\mu_\nu$ is a tensor that projects orthogonally to the target
four-velocity and is given by
\begin{equation}
h^\mu_\nu=-\delta^\mu_\nu-u^\mu u_\nu\ .
\end{equation}
From the above, we see that in order to calculate the radiation
force, we need to calculate its stress-energy tensor $T^{\kappa
\nu}$ in Boyer-Lindquist coordinates, or equivalently
$T^{\hat{\kappa} \hat{\nu}}$ in the LNRF. The two are connected
through the transformation
\begin{equation}
T^{\mu \nu}=e^\mu_{\hat{\kappa}} e^\nu_{\hat{\lambda}}
T^{\hat{\kappa} \hat{\lambda}}\ .
\end{equation}

\subsection{The radiation stress-energy tensor}

\begin{figure}[t]
\centering
\includegraphics[width=0.5\textwidth]{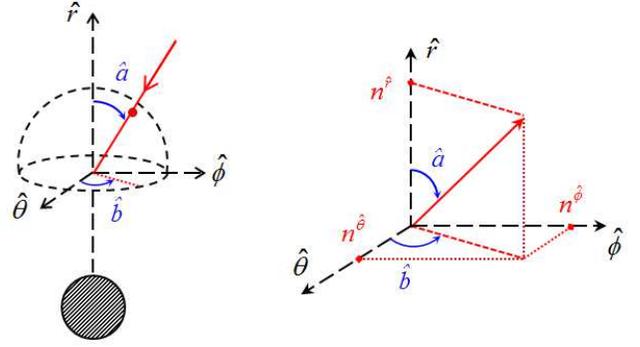}
\caption{The local sky around the target particle. For the direction
of the incoming photon, we define angles $\hat a$ and $\hat b$
similar to the polar angle $\theta$ and the azimuthal angle $\phi$
of the typical spherical coordinates. The striped disk represents
the event horizon of the central black hole.\\}
\label{fig1}
\end{figure}

In the LNRF, the radiation stress-energy tensor is given by
\begin{eqnarray}
T^{\hat{\mu} \hat{\nu}} & = & \int\!\!\int
I_{\nu}(r,\theta,\hat{a},\hat{b};\nu)\ {\rm d}\nu\ n^{\hat{\mu}}
n^{\hat{\nu}}\  {\rm d}\Omega \ ,\nonumber\\
& = & \int I(r,\theta,\hat{a},\hat{b})\ n^{\hat{\mu}}
n^{\hat{\nu}}\  {\rm d}\Omega\ , \label{SET}
\end{eqnarray}
where, $I_{\nu}$ and $I$ are the frequency dependent and
integrated specific intensities respectively in the LNRF at the
position of the target electron. Here, $n^{\hat{\mu}}\equiv
p^{\hat{\mu}}/p^{\hat{t}}$ is the unit spacelike vector along the
photon trajectory that hits the target, and $p^{\hat{\mu}}$ are
the photon four-momentum components in the LNRF. We define angles
$\hat{a}$, $\hat{b}$ as shown in Figure \ref{fig1} 
and therefore we have
\begin{equation}
n^{\hat{r}} = \cos \hat{a},\ n^{\hat{\theta}}
=\sin \hat{a} \cos \hat{b},\  n^{\hat{\phi}} =\sin \hat{a} \sin
\hat{b}\ .
\end{equation}
Radiation photons originate on the surface of the accretion disk,
travel to the position of the target, and reach it from a certain
direction $n^{\hat{i}}$ that corresponds to a solid angle element
${\rm d}\Omega=\sin \hat{a}\ {\rm d}\hat{a}\ {\rm d}\hat{b}$. In
other words, the integral in eq.~(\ref{SET}) has contributions
{\em only} from those directions that correspond to photon
trajectories that originate on the radiation source, in our case
the surface of the hot innermost accretion disk. Therefore, the
calculation of the radiation field requires the {\em backward
integration} of photon trajectories from the position of the
target to their origin on the surface of the disk along all
directions $(\hat{a},\hat{b})$ in the sky of the target particle.

For each such photon trajectory, the specific intensity $I_{\nu}$
that appears in eq.~(\ref{SET}) is {\em different} from the source
specific intensity $I_{\nu, {\rm s}}$. In order to obtain
$I_{\nu}$ we take advantage of the fact that, along the path of a
light ray, $I_{\nu}/\nu^3=I_{\nu, {\rm s}}/\nu^3_{{\rm
s}}$ for the frequency dependent specific intensity, or
equivalently
\begin{equation}
I=\left(\frac{\nu} {\nu_{{\rm s}}}\right)^4 I_{{\rm s}}
\label{Iform}
\end{equation}
for the frequency integrated intensities, where the subscript $s$
refers to quantities calculated at the radius of the source.
Notice that the ratio $\nu/\nu_{\rm s}$ expresses a frequency
shift that is {\em independent} of the frequency itself and
depends only on the form of the spacetime and the angle of
emission. It accounts for three different phenomena: the
gravitational redshift caused by gravitational time dilation, the
Doppler shift caused by the motion of the emitting surface, and
the frame dragging shift caused by the `differential rotation' of
the spacetime. The former two shifts can be encountered in any
spacetime, while the latter only in rotating spacetimes.

\begin{figure}[t]
\centering
\includegraphics[width=0.5\textwidth]{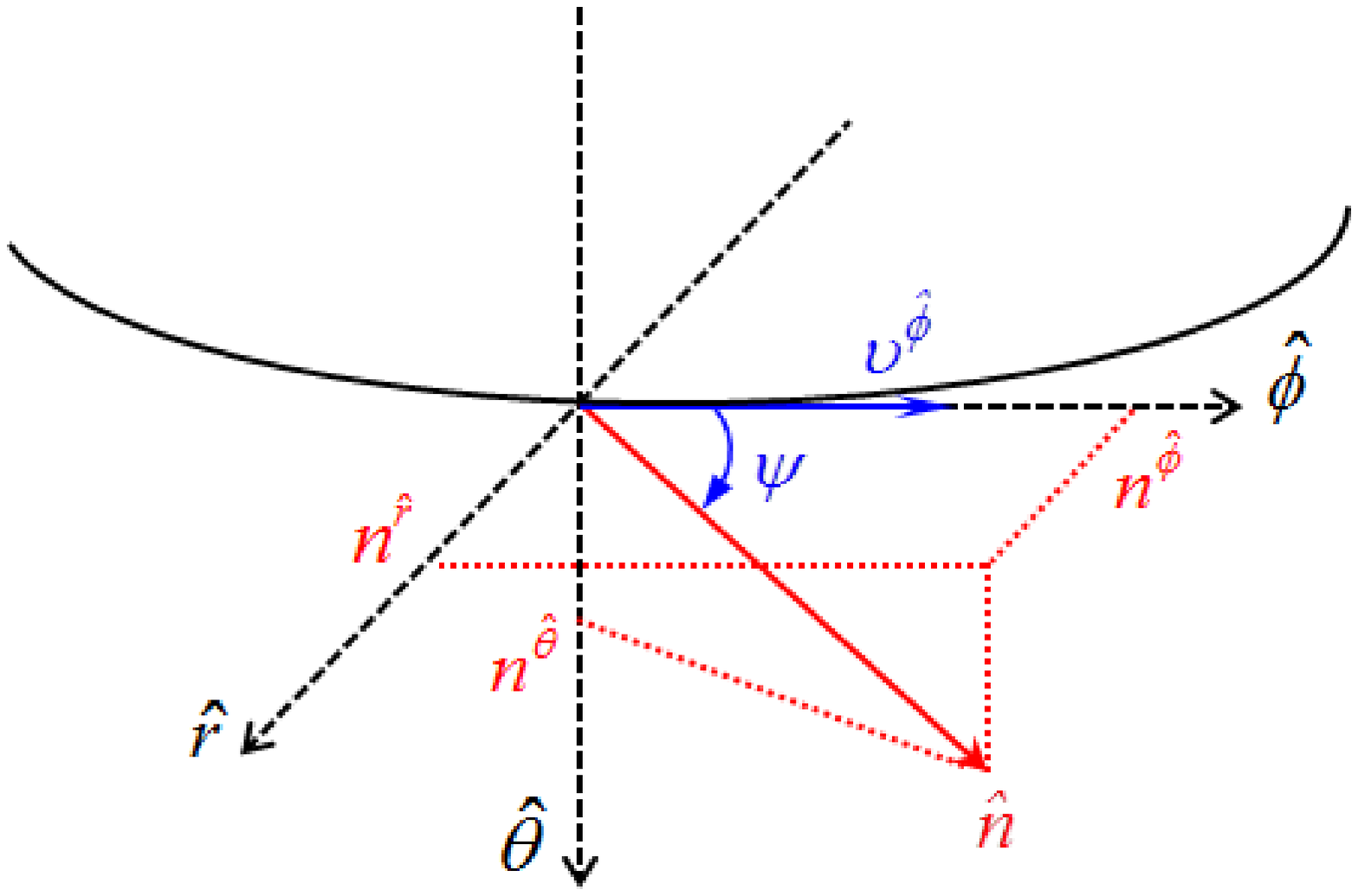}
\caption{The photon emission along the direction $\hat n$ and the
disk motion with velocity $\rm v \equiv \rm v^{\hat \phi}$ at the
point of emission. The arc represents a part of the circular orbit
of the disk element that emitted the photon.} \label{fig6}
\end{figure}

The gravitational redshift between an emitted frequency $\nu'$ and
a received frequency $\nu$ is given by
\begin{equation}
\frac{\nu}{\nu'}=\frac{\alpha'}{\alpha} , \label{gravred}
\end{equation}
which is, as expected, equal to the inverse ratio of the elements
of proper time in each point of the spacetime\footnote{Notice that
eq.~(\ref{gravred}) involves the lapse functions $\alpha$, whereas
the equivalent expression in Miller \& Lamb~1996 involves the metric time
elements $g_{tt}$.}. The Doppler shift can be calculated by the
familiar formula
\begin{equation}
\frac{\nu}{\nu'}=\frac{1}{\gamma(1-{\rm v}\cos\psi)} ,
\label{Doppler}
\end{equation}
where $\gamma=(1-{\rm v}^2)^{-1/2}$ is the Lorentz factor and
$\psi$ is the angle between the direction of motion of the
emitting surface and the direction of the photon emission. Both
$\gamma$ and $\psi$ are measured in the LNRF at the point of
emission (Figure \ref{fig6}). Finally, the frame dragging shift is
equal to
\begin{equation}
\frac{\nu}{\nu'}=\frac{1+\omega\frac{p_\phi}{p_t}}{1+\omega'\frac{p_\phi}{p_t}}
, \label{frdr}
\end{equation}
where $p_\mu$ are the emitted photon covariant four-momentum
components. Notice that the ratio $p_\phi/p_t$ depends on the
direction of emission of each such photon at its origin
(see \S~3 below), and is conserved along the photon trajectory.

\begin{figure*}[b]
\centering
\includegraphics[width=17cm]{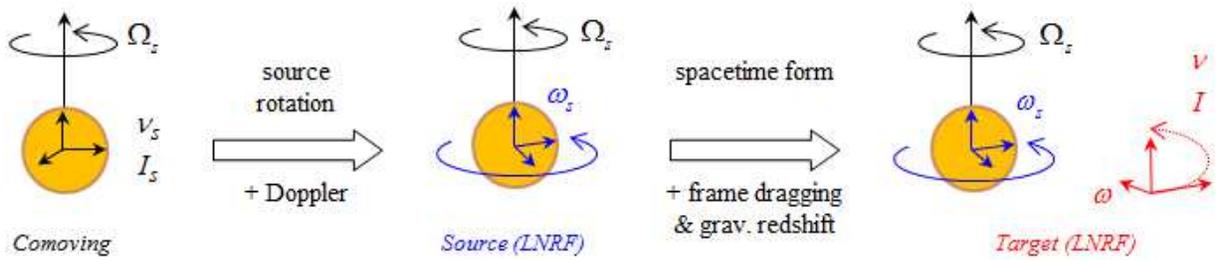}
\caption{Visual representation of the frame change process. When
we are comoving with the source we measure a specific intensity
$I_{\rm s}$. In order to measure it at the LNRF at the source, we
must account for the Doppler shift due to the rotation. To
calculate the specific intensity $I$ observed at the target
electron, we must take into account the spacetime curvature
effects, namely the gravitational redshift due to time dilation
and the frame dragging shift due to the 'differential rotation' of
the spacetime at different radii. The yellow orb is a sphere with
the radius of the emitting surface, whatever that might be (a part
of the accretion disk, the surface of a star, etc) and not the
black hole event horizon.} \label{fig5}
\end{figure*}

In our present study we consider all three of the above shifts
between the frequency $\nu_{\rm s}$ of the emitted photons at
their source as seen by an observer comoving with the source, and
the frequency $\nu$ observed in the LNRF at the position of the
target. The Doppler shift of eq.~(\ref{Doppler}) takes us from the
frame comoving with the source of photons to the LNRF at that
position. Subsequently, the frame dragging shift of
eq.~(\ref{frdr}) takes us from the LNRF rotating with $\omega_{\rm
s}$ at the position of the source to the LNRF rotating with
$\omega$ at the position of the target. In the meantime, the
gravitational redshift of eq.~(\ref{gravred}) accounts for the
difference in time dilation between these two points (see Figure
\ref{fig5}).
Thus, combining eqs.~(\ref{gravred}), (\ref{Doppler}) and
({\ref{frdr}), we deduce the total shift between the emitted and
received frequencies $\nu_{\rm s}$ and $\nu$
\begin{equation}
\frac{\nu}{\nu_{\rm s}}= \frac{\alpha_{\rm s}}{\alpha}~
\frac{1+\omega \frac{p_\phi}{p_t}}{1+\omega_{\rm
s}\frac{p_\phi}{p_t}}~ \frac{1}{\gamma(1-{\rm v^{\hat
\phi}}\cos\psi)} . \label{nutotal}
\end{equation}
From eqs.(\ref{Iform}) and (\ref{nutotal}), the specific intensity
we seek is simply given by
\begin{equation}
I=\frac{\alpha_{\rm s}^4}{\alpha^4}
\left(\frac{1+\omega\frac{p_\phi}{p_t}}{1+ \omega_{\rm
s}\frac{p_\phi}{p_t}}\right)^4 ~\frac{I_{\rm s}}{\gamma^4(1-{\rm
v^{\hat \phi}}\cos\psi)^4}\label{I}
\end{equation}
along each photon trajectory.

\section{Numerical Simulations}

As explained at the end of the previous Section, the complex
astrophysical setup that we are considering (accretion disk +
rotating black hole) requires backward ray tracing along every
direction $(\hat{a}, \hat{b})$ in the sky of the target. In this
present work, we only consider an optically thick accretion disk.
In particular, we only consider target electrons located in a thin
surface layer at the inner edge of the disk at the position of the
ISCO on the equator. The photon trajectories that we obtain are
divided into three categories, depending on their point of origin:
those that cross the event horizon, those that originate from a
point at `infinity' or the `outer disk' (we arbitrarily take
$r>12GM/c^2$) and those that originate from a point of the inner
disk ($r_{\rm ISCO} \leq r \leq 12GM/c^2$). Trajectories from the
first two categories do not give any significant contribution to
the radiation field. Radiation pressure is due to photons of the
third category that originate from the innermost hotter part of
the disk. For those `allowed photon trajectories' we calculate
their point of origin, their direction of emission, and the
resulting frequency shifts between the emission and destination
points. For those photons we will assume that the surface of the
disk radiates as a black body with temperature $T$ that varies as
a known function of distance $r_{{\rm s}}$ on the disk (e.g.
$T\propto r_{\rm s}^{-3/4}$ as in Shakura \& Sunyaev~1976). In
that case, the expression that enters eq.~(\ref{I}) becomes
\begin{equation}
I_{{\rm s}}=\frac{\sigma_B}{\pi} T^4(r_{{\rm s}})
\end{equation}
where $\sigma_B$ is the Stefan-Boltzmann constant (Rybicki \&
Lightman~1986).

In order to implement the above numerical procedure, we developed
two numerical codes:
\begin{enumerate}
\item Code {\bf {\bf \emph{Omega}}} for the ray tracing, and \item
Code {\bf {\bf \emph{Infinity}}} that calculates the
energy-momentum tensor through the integral in eq.~(\ref{SET}) and
the radiation force through eq.~(\ref{force})
\end{enumerate}
Code {\bf \emph{Omega}} calculates single photon trajectories for
given angles $(\hat{a},\hat{b})$ of the incoming photon, finds out
whether the trajectory originates from the inner hotter part of
the disk between $r_{\rm ISCO}$ and $12GM/c^2$ (the outer boundary
is defined arbitrarily for computational covenience), and if yes,
returns its point and angles of origin. All problem parameters can
be easily input by the user. Code {\bf \emph{Infinity}} divides
the sky at the ISCO into half a degree intervals in both $\hat{a}$
and $\hat{b}$, and runs code {\bf \emph{Omega}} for every angle.
This corresponds to roughly 260.000 photon trajectories per run.
It finds the allowed trajectories, calculates the frequency shifts
for each angle, and generates the radiation matrix. From that, the
radiation stress-energy tensor and the four-force components are
computed. We tested our code in the Schwarzschild case with a
central source of radiation and obtained agreement with the
analytic formulae given by Abramowicz, Ellis \& Lanza~(1990) to
within 0.7\% on average.

In Figure~\ref{skymaps}, we present sky maps, namely the sky as
seen in the LNRF at the position of the moving target particle at
the ISCO, for various values of the black hole spin parameter. We
have assumed here that the disk is in Keplerian prograde rotation
around the central black hole. We have also assumed that it is
optically thick. Our calculations were performed with a
Shakura-Sunyaev disk surface temperature profile $T(r_{\rm
s})=10^7\ {\rm K}\ (r_{\rm s}/r_{\rm ISCO})^{-3/4}$. In each
frame, the center of the black hole ($\hat{a}=\pi$) is at the
center, and the direction perpendicular to that ($\hat{a}=\pi/2$)
is on the circumference. The circle of radiation around the black
hole horizon that appears in most of the images is an Einstein
ring generated from the disk upper and lower surfaces. Notice the
difference in the radiation intensity from left to right due to
the material on the right moving toward the target, and the
material on the left moving away from it. Column (a) corresponds
to an infinitesimally thin disk. In this special case, radiation
hits the target even from directions away from the black hole that
correspond to $0\leq \hat{a}<\pi/2$. These directions are not
shown in Figure~\ref{skymaps}, but are taken into account in the
calculation of the integral in eq.~(\ref{SET}). Column (b)
corresponds to a model for a thin disk with height equal to $0.1
M$. Column (c) corresponds to a model for a thick disk with height
equal to $0.5M$. Finally, column (d) corresponds to a thick torus
with circular cross-section that extends from $r=r_{\rm ISCO}$ to
$r=3r_{\rm ISCO}$ along the equator.
\begin{figure*}[h]
\centering
\includegraphics[width=17cm]{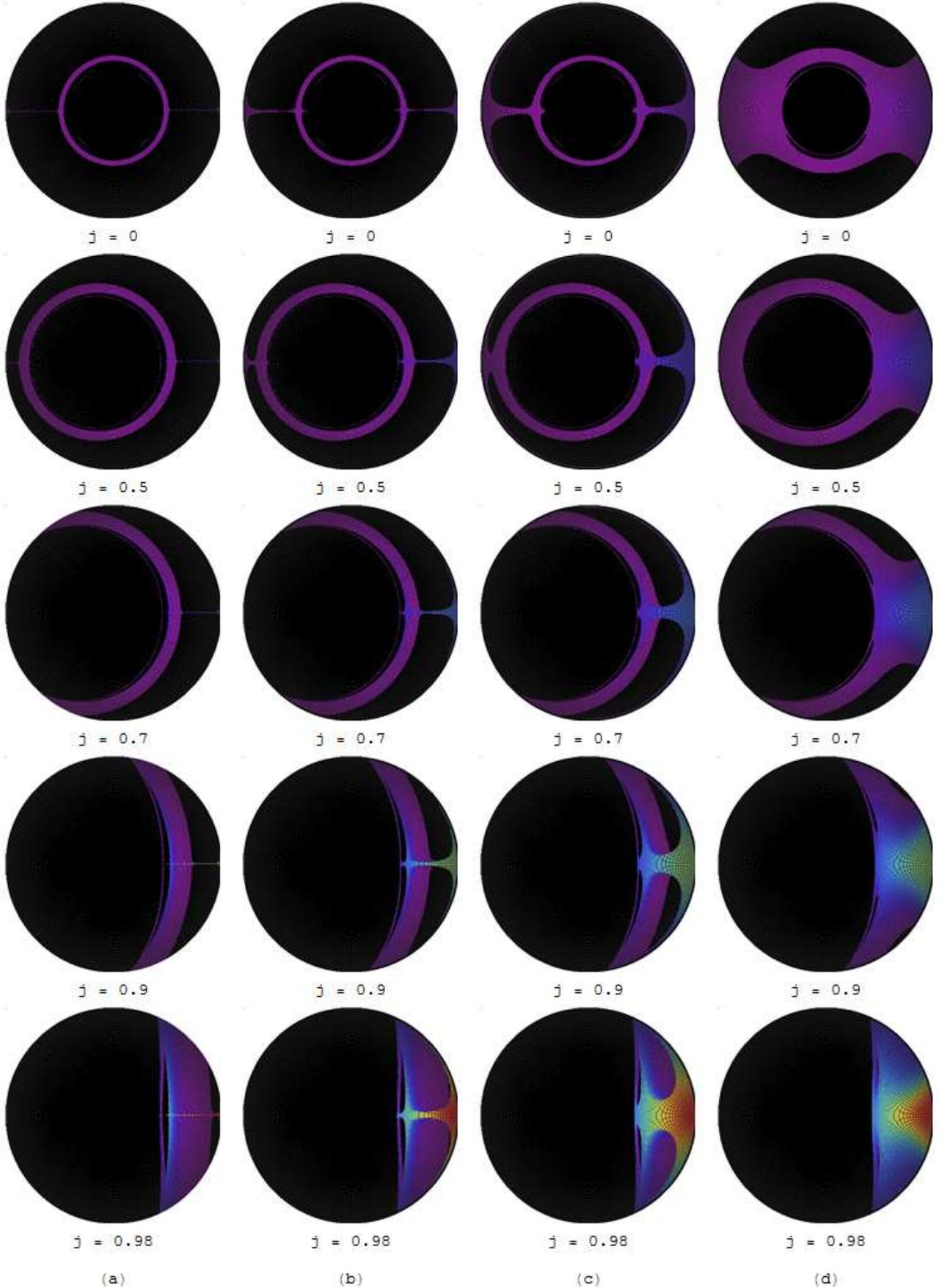}
\caption{The sky as seen in the LNRF at the position of the moving
target particle at the ISCO, for various values of the black hole
dimensionless spin parameter $j\equiv a/M$. The color scale
corresponds to $I(\hat{a},\hat{b})$ at the ISCO. In each frame,
the center of the black hole is at the center ($\hat{a}=\pi$), and
the direction perpendicular to that ($\hat{a}=\pi/2$) is on the
circumference. Column (a): infinitesimally thin disk; column (b):
thin disk; column (c): thick disk; column (d): thick torus (see
text for details). The black hole horizon is the distorted black
circle. The circle of radiation around the black hole horizon is
the Einstein ring from the upper and lower surfaces of the disk.
Notice the difference in radiation intensity from left to right
due to the disk and spacetime rotations. We only considered here
prograde Keplerian accretion disks and a Shakura-Sunyaev surface
temperature profile.} \label{skymaps}
\end{figure*}
For a better understanding of the
rotating spacetime effects on the photon trajectories, the reader
is referred to Appendix B and Figure \ref{eqor} where we show how
the morphology of photon trajectories changes with impact angle
and black hole spin.

In Tables~\ref{tab1a} and \ref{tab1b}, we present our numerical
results for the $r$- and $\phi$-components of the radiation force
per proton respectively (normalized to the canonical gravitational
force per proton $GMm_{\rm p}/r_{\rm ISCO}^2$) for various black
hole spin parameters and for various accretion disk geometries. As
we said before, we have assumed here for simplicity that the
accretion disk is in Keplerian prograde rotation (Bardeen {\em et
al.}~1972) with ${\rm v}^{\hat{r}}\ll {\rm v}^{\hat{\phi}}$ (see
Appendix A). More realistic accretion disk configurations will be
considered in a future work. What we found is most interesting and
rather unexpected:
\begin{enumerate} \item In the theoretical case of an infinitely
thin disk, radiation hits the electrons at the ISCO from all
directions. It is thus expected that the radiation field is almost
isotropic, and therefore, the abberation due to the orbital motion
of the target electrons at the ISCO results in radiation drag.
This is the general relativistic generalization of the PR drag for
thin astrophysical accretion disks. A similar effect is expected
to take place when the accretion disk is geometrically thick but
optically thin, as is the case of an ADAF disk (Narayan \&
Yi~1994). \item As the inner edge of the disk thickens, photons
can reach the ISCO electrons only from the half space facing the
black hole (we have assumed zero optical depth in the disk
interior). The important new element is that the radiation that
originates on the ISCO contributes more and more to the total
radiation pressure, and the radiation field is distorted by the
disk and spacetime rotations. As a result, the radiation force
acts {\em along the direction of rotation}, and the azimuthal
force component changes sign from negative to positive. In other
words, {\em the azimuthal effect of radiation changes from drag to
acceleration}. Considering the analogy with the classical
Poynting-Robertson effect, this result was rather unexpected.
\end{enumerate}
We will next discuss one interesting consequence of our results,
namely the role of the azimuthal radiation drag or acceleration in
the generation of astrophysical magnetic fields.
\begin{table}[t]
\centering \caption{Normalized radial radiation force $f^r_{\rm
rad}/{\frac{GMm_{\rm p}}{r_{{\rm ISCO}}^2}}$}
\begin{tabular}{lcrccc}
\hline \hline
\\
 \text{j=a/M} & $r_{\text{ISCO}}{\rm (M)}$ & \text{Inf. disk} & \text{Thin disk} & \text{Thick disk} & \text{Torus} \\
\\
 \hline
 0 & 6.000 & 0.017 & 0.021 & 0.038 & 0.140 \\
 0.1 & 5.669 & 0.018 & 0.022 & 0.040 & 0.136 \\
 0.2 & 5.329 & 0.018 & 0.023 & 0.041 & 0.132 \\
 0.3 & 4.979 & 0.018 & 0.023 & 0.042 & 0.128 \\
 0.4 & 4.614 & 0.018 & 0.024 & 0.044 & 0.124 \\
 0.5 & 4.233 & 0.018 & 0.024 & 0.046 & 0.120 \\
 0.6 & 3.829 & 0.018 & 0.024 & 0.047 & 0.115 \\
 0.7 & 3.393 & 0.016 & 0.024 & 0.049 & 0.109 \\
 0.8 & 2.907 & 0.013 & 0.025 & 0.052 & 0.102 \\
 0.9 & 2.321 & 0.004 & 0.025 & 0.055 & 0.090 \\
 0.92 & 2.180 & 0.001 & 0.025 & 0.056 & 0.085 \\
 0.94 & 2.024 & -0.004 & 0.025 & 0.056 & 0.080 \\
 0.96 & 1.843 & -0.013 & 0.025 & 0.055 & 0.073 \\
 0.98 & 1.614 & -0.030 & 0.024 & 0.052 & 0.061 \\
 \hline \\
\end{tabular}
\tablecomments{Numerical results for the $r$-component of the
radiation force per proton $f^r_{\rm rad}$, normalized to the
canonical gravitational force per proton $GMm_{\rm p}/r_{\rm
ISCO}^2$ for various black hole spin parameters and accretion disk
geometries. We assumed Keplerian prograde disk rotation for
simplicity, and a Shakura-Sunyaev disk surface temperature profile
$T(r_{\rm s})=10^7\ {\rm K}\ (r_{\rm s}/r_{\rm ISCO})^{-3/4}$.}
\label{tab1a}
\end{table}
\begin{table}[t]
\centering \caption{Normalized azimuthal radiation force
$f^\phi_{\rm rad}/{\frac{GMm_{\rm p}}{r_{{\rm ISCO}}^2}}$}
\begin{tabular}{lcrrrr}
\hline \hline
\\
 \text{j=a/M} & $r_{\text{ISCO}}{\rm (M)}$ & \text{Inf. disk} & \text{Thin disk} & \text{Thick disk} & \text{Torus} \\
\\
 \hline
 0 & 6.000 & -0.007 & 0.003 & 0.037 & 0.111 \\
 0.1 & 5.669 & -0.008 & 0.003 & 0.041 & 0.122 \\
 0.2 & 5.329 & -0.010 & 0.004 & 0.047 & 0.135 \\
 0.3 & 4.979 & -0.012 & 0.004 & 0.054 & 0.153 \\
 0.4 & 4.614 & -0.014 & 0.005 & 0.066 & 0.178 \\
 0.5 & 4.233 & -0.016 & 0.008 & 0.083 & 0.214 \\
 0.6 & 3.829 & -0.019 & 0.017 & 0.117 & 0.275 \\
 0.7 & 3.393 & -0.020 & 0.053 & 0.194 & 0.382 \\
 0.8 & 2.907 & -0.019 & 0.115 & 0.344 & 0.589 \\
 0.9 & 2.321 & 0.002 & 0.307 & 0.833 & 1.187 \\
 0.92 & 2.180 & 0.016 & 0.405 & 1.075 & 1.459 \\
 0.94 & 2.024 & 0.040 & 0.567 & 1.464 & 1.870 \\
 0.96 & 1.843 & 0.092 & 0.863 & 2.179 & 2.576 \\
 0.98 & 1.614 & 0.242 & 1.638 & 3.979 & 4.171 \\
 \hline
 \\
\end{tabular}
\tablecomments{Same as Table~1 for the $\phi$-component of the
radiation force per proton $f^{\phi}_{\rm rad}$. \\} \label{tab1b}
\end{table}

\section{The Cosmic Battery}

Before we proceed, a short introduction on the Cosmic Battery is
in order here. Energetic astrophysical jets have always been
associated with relatively strong large scale magnetic fields that
extract energy from the rotation of a central black hole and its
surrounding accretion disk (Lovelace~1976, Blanford~1976). This
natural association, however, does not answer the fundamental
question of what is the origin of the large scale magnetic field.

Magnetic fields are naturally generated in the disk deep optically
thick interior through a battery mechanism which focuses mainly on
the misalignment between the directions of thermal pressure and
density gradients (Biermann~1950). These fields are subsequently
amplified through dynamo mechanisms based on various rotational
and convective disk instabilities. Unfortunately, when one
considers the back-reaction of the field on the dynamics of the
disk plasma, dynamo action seems to be very inefficient, and as
the field at small diffusive scales reaches equipartition, its
large-scale component remains several orders of magnitude weaker
than the large scale magnetic field expected at the origin of
astrophysical jets (Vainshtein \& Cattaneo~1992, Zrake \&
MacFadyen~2012).

When the disk is non-diffusive, standard MHD advection can bring
the field in from large distances and thus naturally account for
the origin of the large scale magnetic field in the innermost
region of the accretion disk (as attested by the multitude of
ideal MHD simulations of magnetized astrophysical accretion
disks). When this is not the case, i.e. when the magnetic Prandtl
number -the ratio of flow kinematic viscosity to magnetic
diffusivity- in a real astrophysical disk is smaller than some
critical value of order unity (see Contopoulos \& Kazanas~1998 for
details), the Cosmic Battery provides one possible solution to the
same problem. This is a different battery mechanism that focuses
on the anisotropic radiation pressure in the surface layers of the
disk which, as we will see, naturally generates the large scale
dipolar magnetic field expected at the origin of astrophysical
jets.

Eq.~(\ref{E}) yields the plasma electric field as measured in
Boyer-Lyndquist coordinates. A similar result applies in the LNRF,
namely
\begin{equation}
E^{\hat{i}}=\frac{f^{\hat{i}}_{\rm rad}}{\rm e}\ ,
\end{equation}
\begin{equation}
f^{\hat{i}}_{\rm rad}=\sigma_T F^{\hat{i}}\ \mbox{and}\
F^{\hat{i}}=h^{\hat{i}}_{\hat{\nu}}T^{\hat{\kappa} \hat{\nu}}
u_{\hat{\kappa}}\ , \label{flnrf}
\end{equation}
where, $h^{\hat{\mu}}_{\hat{\nu}}=
-\delta^{\hat{\mu}}_{\hat{\nu}}-u^{\hat{\mu}} u_{\hat{\nu}}$ and
$u^{\hat{\mu}}= e^{\hat{\mu}}_\nu u^\nu$. In Table~\ref{tab2a} we
present our numerical results for the $\hat{\phi}$-component of
the electromotive force per proton calculated in the LNRF (once
again normalized to the canonical gravitational force per proton
$GMm_{\rm p}/r_{\rm ISCO}^2$) for various black hole spin
parameters and accretion disk geometries.
\begin{table}[t]
\centering\caption{Normalized azimuthal radiation force
$f^{\hat{\phi}}_{\rm rad}/{\frac{GMm_{\rm p}}{r_{{\rm ISCO}}^2}}$
in the LNRF}
\begin{tabular}{lccccc}
\hline \hline
\\
 \text{j=a/M} & $r_{\text{ISCO}}{\rm (M)}$ & \text{Inf. disk} & \text{Thin disk} & \text{Thick disk} & \text{Torus} \\
\\
 \hline
 0 & 6.000 & -0.007 & 0.003 & 0.037 & 0.111 \\
 0.1 & 5.669 & -0.008 & 0.003 & 0.040 & 0.118 \\
 0.2 & 5.329 & -0.010 & 0.003 & 0.043 & 0.126 \\
 0.3 & 4.979 & -0.012 & 0.002 & 0.047 & 0.134 \\
 0.4 & 4.614 & -0.015 & 0.001 & 0.052 & 0.145 \\
 0.5 & 4.233 & -0.018 & 0.001 & 0.059 & 0.157 \\
 0.6 & 3.829 & -0.023 & 0.002 & 0.071 & 0.175 \\
 0.7 & 3.393 & -0.028 & 0.021 & 0.098 & 0.198 \\
 0.8 & 2.907 & -0.039 & 0.036 & 0.124 & 0.214 \\
 0.9 & 2.321 & -0.061 & 0.054 & 0.163 & 0.230 \\
 0.92 & 2.180 & -0.070 & 0.060 & 0.174 & 0.232 \\
 0.94 & 2.024 & -0.084 & 0.067 & 0.187 & 0.235 \\
 0.96 & 1.843 & -0.106 & 0.075 & 0.203 & 0.236 \\
 0.98 & 1.614 & -0.158 & 0.089 & 0.227 & 0.233 \\
 \hline \\
\end{tabular}
\tablecomments{Same as Table~1 for the $\hat{\phi}$-component of
the radiation force per proton $f^{\hat{\phi}}_{\rm rad}$ in the
LNRF. This is the Cosmic Battery source term in the induction
equation expressed in that frame.} \label{tab2a}
\end{table}
Knowledge of the electric field as seen by ZAMOs in the LNRF
allows us to obtain the rate of change with global time $t$ of the
total poloidal (meridional) magnetic flux $\Psi_{\rm ISCO}$
accumulated inside the ISCO, namely
\begin{equation}
\frac{\partial\Psi_{\rm ISCO}}{\partial t} = 2\pi\alpha\varpi_{\rm
ISCO} \left([{\bf v}\times {\bf B}]^{\hat{\phi}}
-\frac{f^{\hat{\phi}}_{\rm rad}}{\rm e}+\eta [\nabla\times {\bf
B}]^{\hat{\phi}}\right)\ , \label{induction}
\end{equation}
where
\begin{equation}
\Psi_{\rm ISCO}\equiv \pi {\cal B} \varpi_{\rm ISCO} r_{\rm ISCO}
\end{equation}
(eq.~2.20d Macdonald \& Thorne~1982). Here, ${\rm
v}^{\hat{i}}\equiv u^{\hat{i}}/u^{\hat{t}}$, ${\cal B}$ is some
average value of the axial magnetic field threading the region
inside the inner edge of the accretion disk, and $\eta$ is the
effective disk magnetic diffusivity. We repeat once again that, in
order to focus on the effect of the Cosmic Battery in the surface
layers, we have ignored here for simplicity other non-electric
non-gravitational forces that may be important in other parts of
the accretion disk.

The second term in the r.h.s. of eq.~(\ref{induction}) generates
an axial magnetic field which may be along the direction of the
angular velocity vector in the disk if $f_{\rm rad}^{\hat{\phi}}$
is negative (i.e. if radiation results in an azimuthal drag force
in the LNRF at the inner edge of the disk), or opposite to that if
$f_{\rm rad}^{\hat{\phi}}$ is positive (i.e. if radiation results
in an accelerating azimuthal force in the LNRF at the inner edge
of the disk). The magnetic flux that builds up obviously closes
further out through the accretion disk where $f_{\rm
rad}^{\hat{\phi}}$ drops to zero. The axial magnetic field will be
carried by the accretion flow (first term in the r.h.s. of
eq.~\ref{induction}), assuming ideal MHD conditions around and
inside the ISCO, and the flux accumulated inside the inner edge of
the disk will keep growing. The growth will cease, and the
mechanism will saturate if the flow begins to also carry inward
the return polarity of the magnetic field (Contopoulos \&
Kazanas~1998, Bisnovatyi-Kogan~{\em et al.}~2002). Nevertheless,
the return polarity lies in a region where magnetic diffusivity is
significant and the third term in the r.h.s. of
eq.~(\ref{induction}) dominates over the first one
(van~Ballegooijen 1989; Lubow~{\em et al.}~1994; Lovelace~{\em et
al.}~1994; see also, however, Lovelace~{\em et al.}~2009).
Therefore, the return magnetic field {\em diffuses outward through
the disk}, and the mechanism does not saturate but increases the
magnetic field to equipartition values\footnote{This point was
missed by Bisnovatyi-Kogan~{\em et al.}~(2002).}. This natural
scenario where the innermost part of the accretion disk generates
and holds one polarity of the magnetic field, while the return
polarity diffuses outward through the outer diffusive part of the
disk has been called the {\em Cosmic Battery} (Contopoulos \&
Kazanas~1998). Its physical significance is that, since only the
plasma electrons feel the radiation force, they are the only ones
that slow down or accelerate in the azimuthal direction, gradually
building up a relative velocity between themselves and the
protons. This is equivalent to an azimuthal electric current which
gives rise to the poloidal magnetic field through the ISCO. In
other words, the azimuthal radiation force works most efficiently
(maximum radiation pressure, maximum plasma velocities, and
maximum radiation-plasma motion misalignment) to grow
astrophysically significant magnetic fields of order ${\cal
B}_o\sim10^7$~G (for stellar mass black holes) inside the ISCO
over timescales roughly equal to
\begin{equation}
t_{\rm CB}\sim \frac{{\rm e}{\cal B}_o r_{\rm ISCO}}{\alpha_{\rm
ISCO}f^{\hat{\phi}}_{\rm rad}c}\ . \label{tcb}
\end{equation}
The latter estimate is obtained from dimensional analysis of
eq.~(\ref{induction}). The values of $t_{\rm CB}$ that correspond
to ${\cal B}_o=10^7$~G for a $5M_{\odot}$ black hole for various
black hole spin parameters and for various accretion disk models
are shown in
Table~\ref{tab2b}. The characteristic timescales that we obtain
vary from a few hours (in the case of maximally rotating black
holes) to several days (in the case of slowly rotating ones). We
do acknowledge here that we did not take into consideration the
fact that as soon as the magnetic field approaches its
equipartition value, magnetic stresses will begin to affect the
plasma velocity field germane to its origin. Such stresses are
proportional to the square of the magnetic field, thus, they are
expected to become significant only when the magnetic field grows
above about $30\%$ of its equipartition value. Our growth
timescales $t_{\rm CB}$ must, therefore, be taken as rough order
of magnitude estimates.

\begin{table}[t]
\centering \caption{Cosmic Battery timescales $t_{\rm CB}=
\frac{{\rm e}{\cal B}_o r_{\rm ISCO}}{\alpha_{\rm
ISCO}f^{\hat{\phi}}_{\rm rad}c}$ (in hours)}
\begin{tabular}{lcrrrr}
\hline \hline
\\
 \text{j=a/M} & $r_{\text{ISCO}}{\rm (M)}$ & \text{Inf. disk} & \text{Thin disk} & \text{Thick disk} & \text{Torus} \\
\\
 \hline
 0 & 6.000 & 625 & 1532 & 115 & 39 \\
 0.1 & 5.669 & 440 & 1282 & 92 & 31 \\
 0.2 & 5.329 & 301 & 1130 & 72 & 25 \\
 0.3 & 4.979 & 213 & 1348 & 55 & 19 \\
 0.4 & 4.614 & 143 & 1610 & 40 & 15 \\
 0.5 & 4.233 & 93 & 2796 & 29 & 11 \\
 0.6 & 3.829 & 58 & 537 & 19 & 8 \\
 0.7 & 3.393 & 34 & 46 & 10 & 5 \\
 0.8 & 2.907 & 18 & 19 & 5 & 3 \\
 0.9 & 2.321 & 7 & 8 & 3 & 2 \\
 0.92 & 2.180 & 5 & 6 & 2 & 2 \\
 0.94 & 2.024 & 4 & 5 & 2 & 1 \\
 0.96 & 1.843 & 3 & 4 & 1 & 1 \\
 0.98 & 1.614 & 2 & 3 & 1 & 1 \\
 \hline \\
\end{tabular}
\tablecomments{The timescale $t_{\rm CB}$ required for the
build-up of a magnetic field of ${\cal B}_o=10^7$~G around a
$5M_{\odot}$ black hole for various black hole spin parameters and
accretion disk models.} \label{tab2b}
\end{table}

\section{Conclusions}

Our analysis of the radiation field in the immediate vicinity of
an accreting rotating black hole generalizes the classical PR drag
effect in new and unexpected ways. In the case of optically thick
disks, the rotation of the disk and spacetime change the azimuthal
force from drag to acceleration, while the drag is recovered in
the case of infinitely thin disks. It is estimated that optically
thin disks will also experience radiation drag, but this remains
to be shown in future calculations. Moreover, the complex
radiation field is expected to dramatically modify the dynamics of
the disk around its inner edge. Not only that. The azimuthal
component of the radiation force introduces an azimuthal
electromotive source term in the induction equation
which very naturally accounts for the
generation of equipartition-level magnetic fields within
astrophysically relevant timescales (hours to days in stellar
mass black hole X-ray binaries). Note that our present realistic
results differ significantly from previous estimates based on the
classical PR drag effect obtained under the assumption of a
central isotropic luminosity source (Contopoulos \& Kazanas~1998,
Kylafis~{\em et al.}~2012).

We have just began to investigate the astrophysical implications of the
complex radiation field generated by the accretion disk in the
vicinity of a rotating black hole. Our numerical setup is
ideally suited to address several other important effects in
future investigations:
\begin{enumerate}
\item The optical depth of the disk: Photons are emitted and
absorbed from a certain depth below the surface of the disk. This
effect is expected to be very important in optically thin
geometrically thick accretion disks (e.g. ADAF), and in particular
in situations where the type of the inner disk transitions from
optically thin geometrically thick to optically thick
geometrically thin, as is the case in the various stages in the
q-diagram of flaring X-ray binaries (e.g. Belloni~2010).
Consideration of the optical depth will yield a more realistic
distribution of the azimuthal electromotive source of the Cosmic
Battery mechanism. \item A more realistic accretion disk velocity
field where the assumptions that ${\rm v}^r\ll {\rm v}^\phi$ and
${\rm v}(r,\theta)={\rm v}(r,\pi/2)$ in the disk are relaxed.
\item Retrograde disk rotation (our present results were obtained
only for prograde Keplerian rotation). \item More general
temperature profiles. \item The dynamics of the inner accretion
flow: The radial radiation pressure will displace the inner edge
of the disk. At the same time, the azimuthal radiation force is
expected to dramatically modify the accretion flow. Both effects
may be associated with the observed intense variability in X-ray
binaries, disk instabilities, the q-diagram, etc. (Kylafis {\em et
al.}~2012).
\end{enumerate}

We conclude that the radiation field in the vicinity of an
accreting astrophysical black hole modifies the accretion flow and
generates complex electromagnetic dynamics that may be associated
with the origin and evolution of magnetic fields and outflows in
energetic astrophysical sources such as X-ray binaries and AGNs.

\acknowledgements
This work was supported by the General
Secretariat for Research and Technology of Greece and the European
Social Fund in the framework of Action `Excellence'.

\bibliographystyle{apj}
\bibliography{KC2014}

\nocite{*}

\section*{Appendix~A: Particle trajectories}

In the study of particle trajectories in Classical Mechanics, we
often use conserved quantities. In the same manner, when studying
particle trajectories in Kerr spacetime it is not only useful, but
unavoidable to seek and use the integrals of motion, most of which
are already known from Classical Mechanics. A particle with
four-momentum $p=(p^t,p^r,p^\theta,p^\phi)$ in geodesic motion
around a rotating black hole has four conserved quantities: its
rest mass $m$, its total energy $E=-p_t$, its angular momentum
component parallel to the symmetry axis $L=p_\phi$, and the Carter
constant
$Q=p_\theta^2+\cos^2\theta[a^2(m^2-p_t^2)+p_\phi^2/\sin^2\theta]$.
The latter can be thought of as a measure of how much the
trajectory deviates from the equatorial plane (a particle that
starts from the equatorial plane with $Q=0$ will remain there,
whilst a particle that starts outside the equatorial plane with
$Q>0$, will eventually cross it at some point.

Particle trajectories satisfy the equations of motion
\citep[]{B72}
\begin{eqnarray}
\Sigma\frac{{\rm d}r}{{\rm d}\lambda} & = &
\pm(V_r)^{1/2}\nonumber\\
\Sigma\frac{{\rm d}\theta}{{\rm d}\lambda} & = &
\pm(V_\theta)^{1/2}\nonumber\\
\Sigma\frac{{\rm d}\phi}{{\rm d}\lambda} & = &
-( a E-L/\sin^2\theta)+a T /\Delta\nonumber\\
\Sigma\frac{{\rm d}t}{{\rm d}\lambda} & = & -a(aE\sin^2\theta
-L)+(r^2+a^2)T/\Delta \label{eqm1}
\end{eqnarray}
where $\lambda$ is an affine parameter along the trajectory
($\lambda=\tau/m$ for massive particles), and
\begin{eqnarray}
T & \equiv & E(r^2+a^2)-L a\nonumber\\
V_r & \equiv & T^2-\Delta[mu^2 r^2 + (L- a E)^2+Q]\nonumber\\
V_\theta & \equiv & Q-\cos^2\theta [a^2
(m^2-E^2)+L^2/\sin^2\theta]\ .
\end{eqnarray}
For a massive particle like an electron, we can set $m=1$, and for
a massless one like a photon, we take $m=0$. Even though the above
form of the equations of motion is compact and elegant, during
numerical integration it proves to be highly problematic. The
presence of the square roots of the potentials at the first two
equations, accumulates errors rather quickly at the turning
points. Also, in this form, one would have to change the signs of
the square roots by hand at the turning points, something that
causes the execution of even medium resolution codes to be almost
impossible. There are various ways of dealing with the
aforementioned problems. We have found it to be more convenient to
transform the above four equations of motion into an equivalent
Hamiltonian system of six equations (plus two conserved
quantities) and integrate that instead. The new equations of
motion are

\begin{eqnarray}
\frac{{\rm d}t}{{\rm d}\lambda} & = & \frac{1}{2\Sigma\Delta}
\frac{\partial}{\partial E}(V_r+\Delta V_\theta)\nonumber\\
\frac{{\rm d}\phi}{{\rm d}\lambda} & = & -\frac{1}{2\Sigma\Delta}
\frac{\partial}{\partial L}(V_r+\Delta V_\theta)\nonumber\\
\frac{{\rm d}r}{{\rm d}\lambda} & = & \frac{\Delta}{\Sigma}
p_r\nonumber\\
\frac{{\rm d}\theta}{{\rm d}\lambda} & = & \frac{1}{\Sigma}
p_\theta\nonumber\\
\frac{{\rm d}p_t}{{\rm d}\lambda} & = & 0\nonumber\\
\frac{{\rm d}p_\phi}{{\rm d}\lambda} & = & 0\nonumber\\
\frac{{\rm d}p_r}{{\rm d}\lambda} & = &
-p_r^2\frac{\partial}{\partial
r}\left(\frac{\Delta}{2\Sigma}\right)
-p_\theta^2\frac{\partial}{\partial
r}\left(\frac{1}{2\Sigma}\right)\nonumber
+\frac{\partial}{\partial r}\left(\frac{V_r+\Delta V_\theta}
{2\Sigma\Delta}\right)\nonumber\\
\frac{{\rm d}p_\theta}{{\rm d}\lambda} & = &
-p_r^2\frac{\partial}{\partial
\theta}\left(\frac{\Delta}{2\Sigma}\right)
-p_\theta^2\frac{\partial}{\partial
\theta}\left(\frac{1}{2\Sigma}\right)\nonumber +
\frac{\partial}{\partial \theta}\left(\frac{V_r+\Delta
V_\theta} {2\Sigma\Delta}\right)\ . \label{eqm2}
\end{eqnarray}
These are the equations that code {\bf \emph{Omega}} solves
numerically, backtracking the path followed by the photon that hit
the target, until it finds its point of origin on the radiating
disk.

Let us now address another subject that directly relates to the
equations of motion. As we saw in eqs.~(\ref{Doppler}) and
(\ref{flnrf}), in order to calculate the Doppler shift and the
flux in the LNRF at the position of the target, we need to know
the source and target four-velocities. Since both are on the
accretion disk, all we need to calculate is the disk four-velocity
at different radii. Even though our disk models are in general
`thick' (i.e. they extend beyond the equatorial plane), in order
to simplify the process of solution we assume that ${\rm
v}^{\phi}(r,\theta)={\rm v}^{\phi}(r,\pi/2)$. We also consider
only cases with ${\rm v}^r\ll{\rm v}^\phi$, i.e. $\rm v \approx
{\rm v^\phi}$. Solving now the equations of motion (\ref{eqm1})
for circular prograde equatorial orbits, the four-velocity
components are
\begin{eqnarray}
u_t=-\frac{r^{3/2}-2Mr^{1/2}+aM^{1/2}}{r^{3/4}(r^{3/2}-3Mr^{1/2}+2aM^{1/2})^{1/2}} \nonumber\\
u_\phi=\frac{M^{1/2}(r^{2}-2aM^{1/2}r^{1/2}+a^{2})}{r^{3/4}(r^{3/2}-3Mr^{1/2}+2aM^{1/2})^{1/2}}
\label{4u}
\end{eqnarray}
We define the quantity
\begin{equation}
\Omega=\frac{d\phi}{dt}=\frac{u^\phi}{u^t}=\frac{M^{1/2}}{r^{3/2}+aM^{1/2}}\ .
\end{equation}
In order to calculate the velocity needed for the Doppler shift,
simple calculations yield
\begin{equation}
{\rm v^{\hat \phi}}=\frac{u^{\hat \phi}}{u^{\hat t}}=
\frac{\varpi}{\alpha}(\Omega-\omega)\ .
\label{vphihat}
\end{equation}

We can now rewrite eq.~(\ref{nutotal}) explicitly for the frequency shift
%
\begin{equation}
\frac{\nu}{\nu_{\rm s}}= \left(\frac{\Delta_{\rm s} \Sigma_{\rm
s}/A_{\rm s}}{\Delta \Sigma/A}\right)^{1/2}
\frac{1+\frac{2Mar}{A}\frac{p_\phi}{p_t}}{1+\frac{2Mar_s}{A_{\rm
s}}\frac{p_\phi}{p_t}} ~\frac{[1-{\rm (v^{\hat
\phi})^2}]^{1/2}}{1-{\rm v^{\hat \phi}}\cos\psi} \label{nuexp}
\end{equation}
where the subscript $s$ refers to quantities calculated at the radius
of the source. The specific intensity
perceived in the LNRF at the position of the target will hence be
\begin{equation}
I= \left(\frac{\Delta_{\rm s} \Sigma_{\rm s}/A_{\rm s}}{\Delta
\Sigma/A}\right)^{2}
\left(\frac{1+\frac{2Mar}{A}\frac{p_\phi}{p_t}}{1+\frac{2Mar_s}{A_{\rm
s}}\frac{p_\phi}{p_t}}\right)^4 ~\frac{[1-{\rm (v^{\hat
\phi})^2}]^{2}}{(1-{\rm v^{\hat \phi}}\cos\psi)^4}~I_{\rm s}\ .
\label{Iexp}
\end{equation}

In future work, we intend to improve our results by numerically
generating full maps that give the exact velocity and
four-velocity components for every point of the disk. Although the
results will be more accurate, we do not expect them to differ
greatly from the ones presented in Table~1.

\section*{Appendix~B: Numerical codes and photon trajectories}

In this section we describe the main important aspects of our two
numerical codes, {\bf \emph{Omega}} and {\bf \emph{Infinity}}, and
we list their specifications. We also show diagrams generated with
code {\bf \emph{Omega}} that reveal the importance of the
spacetime rotation and its impact on the 'allowed photon
trajectories'.

Code {\bf \emph{Omega}} is responsible for the ray tracing process
and the graphical representation of the compact object, the
accretion disk and the photon trajectories. The code draws the
event horizon, the ergosphere, and the photon sphere. For each
photon trajectory, it displays its length and end (at the target
electron), along with a marker at the source whose color
represents the type of point of origin. For the accretion disk,
the inner, hotter part is drawn, along with the ISCO and the outer
radius outside which we consider the disk to be cold and thus
emitting negligible radiation. If the user desires so, an
extension of the cold disk can be drawn for representation.
Finally, for every trajectory the three constants of motion $E$,
$L$ and $Q$ are shown along with the photon four-momentum product
$p_\mu p^\mu$ at the beginning and the of the orbit (or any other
intermediate point) in order to ensure that it is close to zero,
as it should be for a photon. The code at the end returns and
outputs the coordinates of the point of origin.

The interface of code {\bf \emph{Omega}} is user-friendly with
buttons, sliders or input boxes that allow total command on any
aspect of the problem. The controllable fields are
\begin{enumerate}
\item {\bf The disk model:} The user can choose the shape of the
disk from the following:
\begin{enumerate} \item no accretion disk,
\item a zero-thickness band of radiation at the distance of the
ISCO, \item an infinitely thin disk at the equatorial plane, \item
a slab of equal thickness at all points with half-height $\rm h$,
\item a wedge-like accretion disk of equal angle at all points and
a half-height $\rm h$ at the ISCO and \item a torus whose
generating circle has a radius $r_{tor}$ and its inner edge lies
on the ISCO.
\end{enumerate} \item {\bf Cold disk:} The user can choose whether
a part of the cold disk is visualized. \item {\bf The upper limit
for the affine parameter $\lambda$:} The trajectory integration
normally stops at an arbitrary radius of $12M$ but the user can
choose to stop it at any point before or after that, as long as no
point of origin is found. \item {\bf The spin parameter:} The
dimensionless spin parameter $j=a/M$ is set. \item {\bf The target
electron position:} By default, the target electron is set at the
equatorial plane ($\theta=\pi/2$) at the distance of the ISCO
($r=r_{ISCO}$). It is possible though for the user to input
initial positions of his/her own choice. \item {\bf The band
half-height:} The user can control the height of the disk at the
distance of the ISCO for the models that have the corresponding
dimension. \item {\bf The angles of photon incidence:} The angles
$\hat a$ and $\hat b$ of the incoming photon can be input by the
user with any accuracy. The former takes values from $0^\circ$ to
$180^\circ$ and the latter from $0^\circ$ to $360^\circ$. If the
user however desires to enter different values, this is also
possible. \item {\bf The zoom:} The user can zoom in or out of the
picture in order to view the result with more detail or to see the
bigger picture. The output 3D image of the disk and the photon
trajectory can also be rotated in any direction to allow better
viewing for the various trajectories.
\end{enumerate}

Code {\bf \emph{Infinity}} first separates the sky of the target
particle into a grid. Then, it uses code {\bf \emph{Omega}} as a
subroutine and scans across that grid in order to calculate the
frequency integrated specific intensity at each point. The various
shifts for each trajectory are calculated by separate subroutines.
After the radiation matrices are created, the code calculates the
stress-energy tensor in the LNRF using eq.~(\ref{SET}). Then it
calculates the radiation pressure force through eqs.~(\ref{force})
or (\ref{flnrf}), depending on the user's choice of reference
frame. The interface of code {\bf \emph{Infinity}} is also
user-friendly and compact. It is composed of seven popup menus
that allow the user to fully specify the parameters of the
problem. These fields are
\begin{enumerate}
\item {\bf The disk model:} Similar to code {\bf \emph{Omega}}.
The user can choose between the band, the thin disk, the slab, the
wedge and the torus. \item {\bf The disk temperature model:} The
user chooses between an isothermal disk with $T(r_{\rm
s})=$~const., or an accretion disk with $T(r_{\rm
s})\propto r_{\rm s}^{-3/4}$. \item {\bf Grid size:} The
resolution of the target's sky grid is set by the user. Although
in our study we work with a 0.5$^\circ$ x 0.5$^\circ$ grid,
depending on the the time and accuracy needs, one can choose from
a variety of grid sizes ranging from 10$^\circ$ x 10$^\circ$ all
the way down to 0.25$^\circ$ x 0.25$^\circ$. \item {\bf Disk
half-height:} Similar to code {\bf \emph{Omega}}. It controls the
height of the disk at the distance of the ISCO for the band, the
slab, the wedge and the torus models. \item {\bf Black hole mass:}
The central black hole mass can be chosen from a list ranging from
$1M_{\odot}$ up to $4.5\times 10^6M_{\odot}$ (the estimated mass
for a Milky Way-size supermassive black hole). \item {\bf Spin
parameter:} The spin parameter $a$ of the black hole can be chosen
from a list of more than 20 values from 0 to $0.999M$. \item {\bf
Reference frame:} Simulation results can be expressed either in
the LNRF or the Boyer-Lindquist frame at the position of the
target.
\end{enumerate}

Finally, in Figure~\ref{eqor} we present representative photon
trajectories as obtained using code {\bf \emph{Omega}} in order to
elucidate the main effects introduced by the spacetime rotation.
The images depict the system as seen from above the equatorial
plane. The changes in the orbits are easily noticeable as the
black hole spin increases.

\begin{figure*}[t]
\centering
\includegraphics[width=17cm]{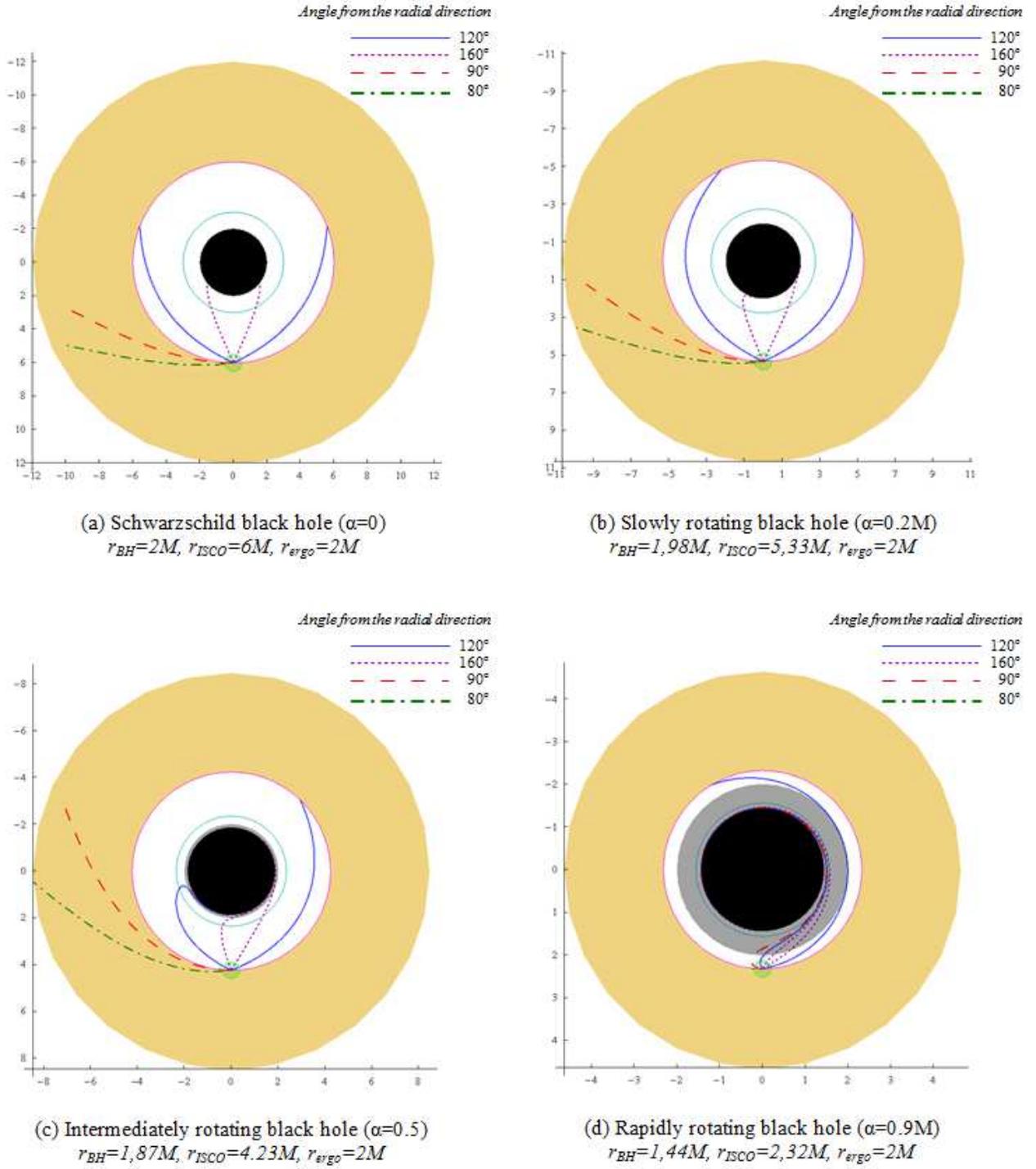}
\caption{The equatorial plane and photon trajectories around
clockwise rotating black holes of different spin parameters. The
black disk is the event horizon, the grey annulus is the
ergosphere, the inner circle is the photon sphere, the outer is
the ISCO and the shaded area outside it, is the rest of the
accretion disk. The trajectory angles are measured from the
(outward) radial direction and the accretion disk rotates
clockwise. Notice that as $a$ increases, we zoom into the picture
in order to better see the details. The effects of the rotation on
the photon trajectories are clearly visible as the spin parameter
increases. The allowed photon trajectories tend to concentrate in
the direction of motion.} \label{eqor}
\end{figure*}

\end{document}